\documentclass[namedreferences]{solarphysics}

\usepackage[optionalrh]{spr-sola-addons} 
\usepackage{graphicx}        

\newcommand{\aap}{    {\it Astron. Astrophys.}}

\newcommand{\apj}{    {\it Astrophys. J.}}
\newcommand{\apjl}{   {\it Astrophys. J. Lett.}}

\newcommand{\mnras}{  {\it Mon. Not. Roy. Astron. Soc.}}

\newcommand{\solphys}{{\it Solar Phys.}}
 
\newcommand{\ssr}{    {\it Space Sci. Rev.}}

\begin{document}

\begin{article}
\begin{opening}

\title{A Relationship between the Solar Rotation and Activity Analysed by Tracing Sunspot Groups}

\author[addressref={aff1},corref,email={rdomagoj@geof.hr}]{\inits{D.}\fnm{Domagoj}~\lnm{Ru\v{z}djak}\orcid{0000-0002-4861-0225}}
\author[addressref={aff1}]{\inits{R.}\fnm{Roman}~\lnm{Braj\v{s}a}}
\author[addressref={aff1}]{\inits{D.}\fnm{Davor}~\lnm{Sudar}}
\author[addressref={aff1}]{\inits{I.}\fnm{Ivica}~\lnm{Skoki\'c}}
\author[addressref={aff2}]{\inits{I.}\fnm{Ivana}~\lnm{Poljan\v{c}i\'c Beljan}}

\address[id=aff1]{Hvar Observatory, Faculty of Geodesy, University of Zagreb, Ka\v{c}i\'ceva 26, 10000 Zagreb, Croatia}
\address[id=aff2]{Department of Physics, University of Rijeka, Radmile Matej\v{c}i\'c 2, 51000 Rijeka, Croatia}

\runningauthor{D. Ru\v{z}djak et al.}
\runningtitle{Solar Rotation and Activity}

\begin{abstract}
The sunspot position from Greenwich Photoheliographic Results \newline(GPR),
US Air Force Solar Optical Observing Network and National 
Oceanic and Atmospheric Administration (USAF/NOAA), and 
Debrecen Photoheliographic Data (DPD) data bases in the period 1874 to 2016 
were used to calculate yearly values of the solar differential-rotation parameters $A$ and $B$. 
The calculated differential-rotation parameters were compared with the solar-activity level.
We found that the Sun rotates more differentially at the minimum than at the maximum of activity 
during the 1977\,--\,2016 epoch. An inverse correlation between equatorial rotation and solar activity 
was found using the recently revised sunspot number. The secular decrease of equatorial rotation rate
accompanying the increase of activity stopped in the last part of the 20th century.
It was noted that when a significant peak of equatorial rotation velocity is observed during minimum 
of activity, the strength of the next maximum is smaller than the previous one. 
\end{abstract}
\keywords{ differential rotation $\cdot$ sunspot groups $\cdot$ activity cycle}
\end{opening}

\section{Introduction}

Solar activity manifests itself as many diverse phenomena (sunspots, flares, coronal mass ejections, {\it etc.}) 
which vary on several distinct timescales. The most prominent manifestation of solar activity 
is the 11 year Schwabe cycle. Extensive reviews on the long-term behaviour of the solar activity and the 
Schwabe cycle are given by \citet{hathaway2010} and \citet{usoskin2017}. 

It is generally accepted that the solar cycle is magnetic in nature and generated by dynamo processes 
within the Sun. The Sun's magnetic field is maintained by its interaction with plasma motions, 
{\it i.e.}, differential rotation of the convective zone. 
The physical processes underlying the complex magnetohydrodynamical system are still not 
fully understood and a large number of dynamo models have been considered. 
An extensive review of solar-dynamo models is given by \citet{Cameron2016}. 
Precise measurements of the solar differential rotation, its variations, and its correlation with the solar activity give important observational constraints on the modelling of the solar dynamo.   

The determination of the solar differential rotation has been performed by many authors using various 
methods. Most often sunspots and sunspot groups have been used 
as tracers because they are well-defined structures with sufficient lifetime and long series of
observations are available. Their disadvantages are complex structure and evolution, 
limb-effects and narrow latitudinal distribution. 
Most extensively, the Greenwich Photoheliographic Results (GPR) were used, {\it e.g.}, by \citet{newton1951}, 
\citet{balthasaretal1986},  \citet{brajsaetal2002}, \citet{ruzdjak2005},  
among many others. The GPR data were often extended by US Air Force Solar Optical 
Observing Network and National Oceanographic and Atmospheric Administration (USAF/NOAA) data \citep{pulkkinen1998,zuccarello2003,ruzdjak2004,brajsa2006,sudar2014,javaraiah2016}.
Besides GPR and USAF/NOAA sunspot data, a number of other datasets were also  frequently used, {\it e.g.}, the Mt. Wilson dataset \citep{Howard1984,gilman1984} which consists from the areas 
and positions of the sunspots measured on the white-light photographs
of the Sun taken with the 30-cm diameter lens at the Tower Telescope
at Mt. Wilson Observatory, Kodaikanal data \citep{gupta1999}, Kanzelh\"ohe 
data \citep{lustig1983,poljancic2014,poljancic2017} and Debrecen Photoheliographic Data (DPD) \citep{sudar2017}.
Apart from sunspots and sunspot groups, other data were used for determination of solar differential 
rotation: for instance, Doppler shifts \citep{howard1976}, coronal holes \citep{nash1988},
faculae \citep{meunier1997}, H$\alpha$ 
filaments \citep{brajsa1997}, global magnetic fields \citep{obridko2001}, helioseismology \citep{howe2000,antia2001,komm2017}, solar radio emission \citep{li2012},
coronal bright points (CBP) \citep{woehl2010,sudar2015,sudar2016}, and 
coronal green line emmision \citep{rybak1994,badalyan2016}. 

The inferred solar rotation was often analysed for long-term changes or changes within a cycle
\citep[see, {\it e.g.},][]{howard1976,gilman1984, gupta1999}. More recently \citet{brajsa2006}, using rotation-rate residuals calculated from sunspot groups, found secular deceleration of 
rotation and faster rotation at the minimum than at the maximum of the solar cycle. 
\citet{jurdana2011}, using rotation rates obtained by 
tracing CBP, investigated a relationship between the solar-rotation parameters and solar activity
expressed by the sunspot number, and they found a significant correlation for parameter $A$
and a small, statistically insignificant, negative correlation for differential-rotation parameter $B$.
\citet{li2014} revisited already published solar-rotation data obtained from H$\alpha$ charts
and sunspots and confirmed the secular deceleration and negative correlation between the differential-rotation parameter $A$ and solar activity. 
Most recently, \citet{badalyan2016} analysed the 
coronal green-line data from 1943 up to the present and found that the equatorial rotation rate
increases in the epochs of minimum between the even and odd-numbered solar cycles and reaches its minimum 
between the odd and even cycles, and that the differential gradient increases in the even cycles,
with its largest values near the maximum of the activity cycle.
On the other hand, theoretical calculations \citep{brun2004,brunetal2004,lanza2006,lanza2007} predict that 
the Sun should rotate more rigidly at the maximum than at the minimum of activity.

To gain insight into these discrepancies between the observations and theoretical predictions, 
in this work we examine the relationship between solar activity and differential-rotation 
parameters determined from the sunspot-group position data from GPR, USAF/NOAA, and 
DPD in the period from 1874 to 2016. The sunspot-group position data were used to 
calculate yearly differential-rotation parameters and the obtained parameters are then 
compared with  the solar activity.
Further, during 2011\,--\,2014 a series of four workshops was organized with the goal of providing a 
community-vetted time series of sunspot numbers for use in long-term studies \citep{cliver2013,cliver2015}.
This effort resulted in recalibration of sunspot and group numbers \citep{clette2014,clette2015,clette2016}.
The revised sunspot numbers are used as a measure of  solar activity in this work.


\section{Data and Reduction}

In this work the sunspot-group data from GPR (1874\,--\,1976), USAF/NOAA Sunspot Data,
and DPD \citep{dpd1,dpd2} sunspot database (1977\,--\,2016) were used.  Our own GPR 
digital dataset was used, while USAF/NOAA and DPD data were downloaded from 
their websites (\textsf{solarscience.msfc.nasa.gov/greenwch.shtm} and \textsf{fenyi.solarobs.csfk.\-mta.hu/DPD/}, respectively). 
The given daily positions of the sunspot groups have been used to 
obtain the rotational velocities. For the groups for which the central meridian 
distance (CMD) was less than 58$^\circ$, which corresponds to about 0.85 of 
projected solar radius \citep[see][]{balthasaretal1986}, the rotation 
rate was calculated by division of CMD differences by elapsed time. The obtained 
synodic rotation velocities were then transformed to sidereal ones by the procedure 
described by \citet{rosaetal1995}, \citet{brajsaetal2002} and \citet{skokic2014}.
Finally the sidereal rotation velocities amounting to less than 8$^\circ$\,day$^{-1}$ 
and exceeding 19$^\circ$\,day$^{-1}$ have been regarded as erroneous and have not 
been considered in the analysis. The constraints described left 92,762 rotation velocities 
(out of 161,714 sunspot position data) obtained from GPR, 58,219 rotation velocities 
(out of 106,981) and 48,814 (out of 84,449) for the DPD and USAF/NOAA sunspot databases respectively.

The solar differential rotation is usually represented by
\begin{equation}
\omega=A+B\sin^2\varphi +C\sin^4\varphi
\end{equation}
where $\omega$ is the sidereal angular rotation velocity, $\varphi$ is the heliographic 
latitude and $A$, $B$, $C$ are differential-rotation parameters. 
Because of the latitudinal distribution of sunspots, the last term can be neglected when
using sunspot motions to derive the parameters, 
{\it i.e.}, $C=0$. Differential-rotation parameters $A$ and $B$ were obtained by fitting 
the resulting equation to all points in the data set. The results for individual 
datasets are presented in Table \ref{rotpar}. The parameters $A$ and $B$ were also 
calculated separately for each year in the timespan of the datasets (1874\,--\,2016).
Data from both of the solar hemispheres were treated together to give a statistically 
significant number of data points for the determination of differential-rotation 
coefficients in the years near minimum of activity.

The solar-activity data (total sunspot number) were taken from  
WDC-SILSO (\textsf{sidc.oma.be/silso/}), Royal Observatory of Belgium, Brussels.
The version 2.0 of the data containing a new entirely revised data series, available since 
1 July 2015, was used. The yearly mean total sunspot number, obtained by taking a simple 
arithmetic mean of the daily total sunspot number over all of the days of one year, was 
taken as the measure for the solar activity of a given year.

   \begin{table}
      \caption[]{Differential-rotation parameters for various datasets: 1 - GPR; 2 - USAF/NOAA;
3 - DPD; 4 - GPR+USAF/NOAA; 5 - GPR+DPD.  (*The USAF/NOAA set contains only Jan.-Sept. data for 2016).}
         \label{rotpar}   
         \begin{tabular}{lccc}
            \hline
                       Timespan &$A$ [$^\circ$\,day$^{-1}$]  & $B$ [$^\circ$\,day$^{-1}$] & Dataset   \\
                       \hline
                       1874-1976 & 14.528$\pm$0.006 &-2.77$\pm$0.05 &1\\
            1977\,--\,2016* & 14.44$\pm$0.01 &-2.54$\pm$0.08 &2 \\
            1977\,--\,2016 & 14.403$\pm$0.009 &-2.44$\pm$0.08 &3 \\
            1874\,--\,2016* & 14.501$\pm$0.005 &-2.71$\pm$0.05 &4 \\
            1874\,--\,2016 & 14.483$\pm$0.005 &-2.67$\pm$0.05 &5 \\
                       \hline
         \end{tabular}
   \end{table}

\section{Results}
\subsection{Secular Changes of the Solar Rotation and Activity}
\label{changessec}

\begin{figure}
   \includegraphics[width=\columnwidth]{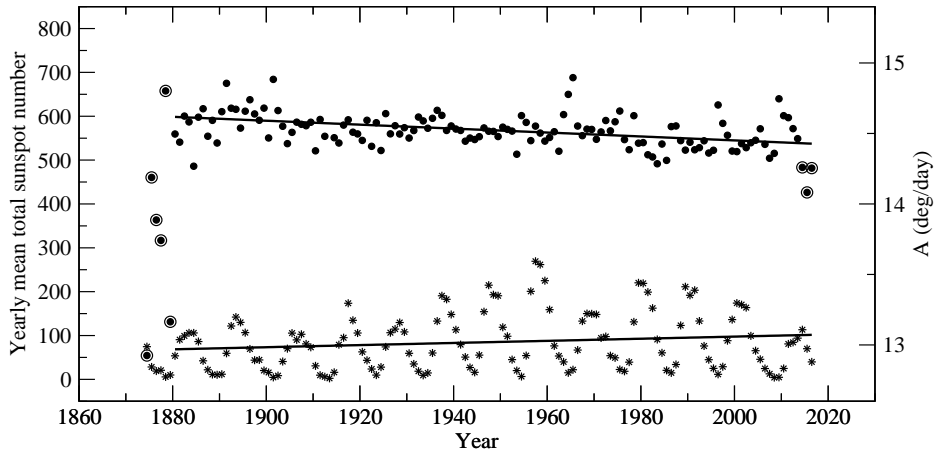}
   \caption{Secular variations of solar rotation and activity. 
The yearly values of equatorial rotation rate $A$ (black circles) 
are shown together with the yearly mean total sunspot number (asterisks).
The lines represent linear fits of the data for the time interval 
(1880\,--\,2016). The GPR and DPD data are shown. The circled points denote 
the suspicious small values of the equatorial rotation at the edges of 
the observing interval
              }
         \label{aintime}
   \end{figure}

For illustration of the secular changes of solar rotation and activity, we present 
in Figure \ref{aintime} the yearly values of the differential-rotation coefficient $A$
(equatorial rotation velocity) together with the yearly mean total sunspot number. 
Rotation rates are denoted with circles and mean sunspot numbers with asterisks. 
The lines represent the linear fits to the data in the 1880 to 2016 period. Only the 
GPR and DPD rotation data are shown. During the last 150 years, on average, solar 
activity was growing, reaching its maximum in the second half of the 20th century 
and has been decreasing since. This can be noted by examining the amplitude (strength) of  
the maxima of Solar Cycles 12\,--\,24, which are represented by the yearly mean total 
sunspot numbers in the lower part of Figure \ref{aintime}. 
This increase/decrease would be  monotonic, but the strength of the Cycle 14, 16, and 20 
maxima are too low to support this interpretation. On the other hand, the solar 
equatorial rotation shows a steady decrease during this period. The exceptions are 
the data for first six years (1874\,--\,1879), whose values (except 1878) are systematically 
lower than the rest of the data. Those data (circled points in the left side 
Figure \ref{aintime}) were excluded from the calculation of the secular trend. 
Linear fits give for the increase of activity 0.2$\pm$0.1 year$^{-1}$ 
and -0.0014$\pm$0.0003$^\circ$\,day$^{-1}$\,year$^{-1}$ between 1880 and 2016 
 and -0.0012$\pm$0.0003$^\circ$\,day$^{-1}$\,year$^{-1}$ for 1880\,--\,2013 for the decrease 
of the equatorial rotation rate.
The obtained value is the same as the values obtained by \citet{brajsa2006} for rotation 
rate residuals and by \citet{li2014} for comparison of previous results for different 
methods and data sets.

It is worth pointing out that the equatorial rotation-velocity minimum in 2014\,--\,2016 
(circled points in the right side of Figure \ref{aintime}) is 
not present in the USAF/NOAA data. In fact the  USAF/NOAA data give a rise of 
equatorial rotation of 0.004$\pm$0.002$^\circ$\,day$^{-1}$\,year$^{-1}$, 
for the 1977\,--\,2016 interval, but it has suspicious small values of the 
coefficient $A$ for the first two years (1977\,--\,1978). 
These results show how difficult it can be to determine the secular trend 
when it is ''obscured" by the cycle-related variations. It is unfortunate 
that all of the datasets have some extreme points near the edges. Besides the
suspiciously low values at the beginning, due to cycle-related variations the GPR dataset
has its maximal yearly value of 14.90$\pm$0.15$^\circ$\,day$^{-1}$ near the end of observing period (1965), which 
also makes the determination of the secular trend more complicated. 
To work around this problem, besides omitting the suspicious points, 
all further analyses are made by taking into account the errors of 
yearly values of solar differential-rotation parameters. 
The results are similar if the ''problematic" points are omitted or 
all data are considered, but with errors taken into account.
Similar deceleration is obtained (-0.0016$\pm$0.0002$^\circ$\,day$^{-1}$\,year$^{-1}$)
for 1874\,--\,2016 when the errors are considered. Individual sets yield: 
-0.0007$\pm$0.0003$^\circ$\,day$^{-1}$\,year$^{-1}$ (GPR 1874\,--\,1976),\linebreak 
-0.0007$\pm$0.0010$^\circ$\,day$^{-1}$year$^{-1}$
(DPD 1977\,--\,2016), and 0.004$\pm$0.001$^\circ$\,day$^{-1}$year$^{-1}$ 
(USAF/NOAA 1977\,--\,2016). 
These results indicate that the secular decrease of the rotation velocity stopped sometime 
in the second half of 20th century.

It was mentioned that the maximal equatorial rotation rate was observed in 1965.
The peak of equatorial rotation is observed during the minimum 
of Solar Cycle 20. The (weighted) mean value of equatorial rotation 
during minimum (14.69$\pm$0.05, 1963\,--\,65) is significantly (2$\sigma$) higher than
the value for the preceding and following years (14.47$\pm$0.03, 1960\,--\,62 and
14.52$\pm$0.03, 1966\,--\,68). The similar value of 14.88$\pm$0.18$^\circ$\,day$^{-1}$ 
was observed during minimum of Solar Cycle 14.
It can be noted in Figure \ref{aintime} that Solar Cycles 
14 and 20 have a smaller amplitude than ''expected" (smaller than what is needed for 
the envelope to be monotonic).  Similar peaks of rotation velocity are visible for Cycles 
23 and 24 but the peak is not observed for Cycle 16 (although there is $\approx$0.25$^\circ$\,day$^{-1}$ 
jump during the previous minimum). Therefore, it can be concluded that if the peak of the 
equatorial rotation velocity is observed during minimum, the following maximum is weaker 
than the previous one.

\begin{figure}
   \includegraphics[width=\textwidth]{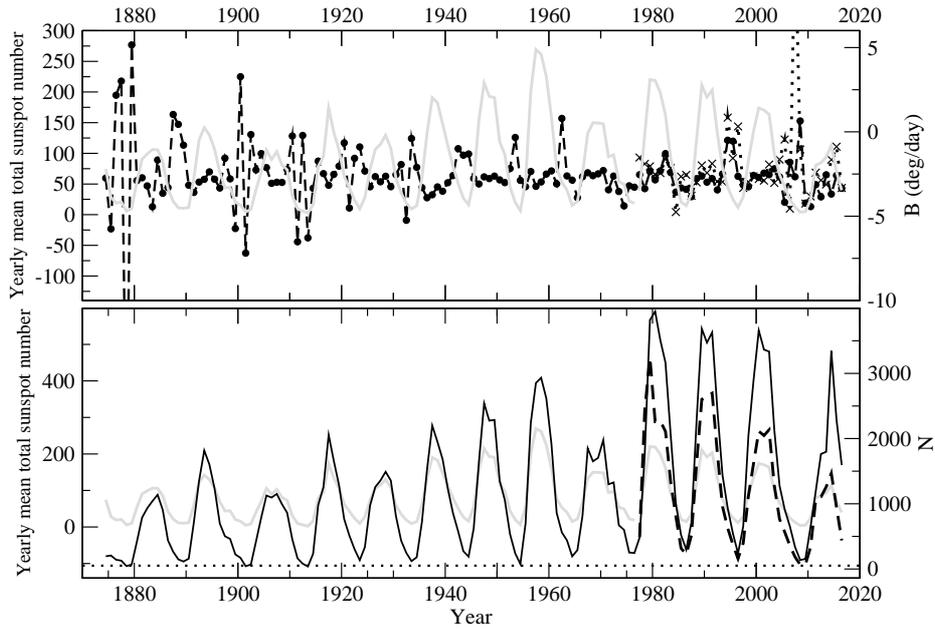}
      \caption{Secular variations of the differential-rotation parameter $B$ 
(upper panel). GPR and DPD data are represented by points connected with the dashed 
line, while USAF/NOAA data are shown with \textsf{x}'s connected by the dotted line. 
In lower panel the number of data available for given year is shown. GPR and DPD 
data are represented with the solid line and USAF/NOAA with the dashed line. 
Horizontal-dotted line represents $N=50$ level. The yearly mean total sunspot
number is represented by the grey line on both panels.
              }
         \label{Bintime}
   \end{figure}

The secular variation of the differential-rotation parameter $B$ is shown in 
upper panel of Figure \ref{Bintime}. 

By inspection of the figure it is hard to see any correlation between the changes 
of solar activity and differential-rotation parameter $B$. During the maximum 
of activity $B$ has values near to its average value, while it oscillates 
during the solar-activity minima. This behaviour is due to the small number of 
data in years near minimum. This is illustrated in the lower panel of the 
figure where the total number of data is shown. It can be seen that in some 
solar minima this number is lower than 50 (dotted-horizontal line in the figure). 
In Table \ref{minmax} the values for differential-rotation parameters for 
several minima and maxima are presented. Inspecting the values in the table 
two conclusions can be drawn. First, the values of equatorial rotation velocity 
are on the average higher at the minimum than at the maximum of activity (weighted means 
are 14.57$\pm$0.04$^\circ$\,day$^{-1}$ and 
14.44$\pm$0.02$^\circ$\,day$^{-1}$, for the minimum and maximum of activity, respectively, which is above the 2$\sigma$ level). 
Second, about 50\,\% of the values for differential-rotation parameter $B$ in the 
years near minimum of activity have large errors. In the analysis of
such a dataset it is vital that errors of individual values are considered.

Finally we would like to draw attention to the differential-rotation 
parameters for individual sets presented in Table \ref{rotpar}. The 
equatorial rotation velocity from GPR is significantly higher than 
the one in USAF/NOAA and DPD data. The difference between USAF/NOAA and 
DPD is (just 0.001$^\circ$\,day$^{-1}$) inside the 2$\sigma$ range. 
The absolute value of the parameter $B$ for the GPR data is 2$\sigma$ larger 
than the value for 
DPD and 1.77$\sigma$ larger than the value for the USAF/NOAA, while DPD and 
USAF/NOAA have values within 1$\sigma$. This means that the equatorial 
rotation rate is on average larger during 1874\--\,1976 than in the 1977\,--\,2016 
period and maybe the Sun rotated more differentially in the earlier epoch. 

  \begin{table}
      \caption[]{Values and errors of differential-rotation coefficients 
[$^\circ$\,day$^{-1}$] for several solar-cycle minimum and maximum years 
covered by the data. (1 - GPR; 2- USAF/NOAA; 3 - DPD).}
         \label{minmax}   
         \begin{tabular}{cccrcccrr}
            \hline
                              &minimum& & & & maximum & & &\\
             Year & $A$ & $B$ & $N$ & Year & $A$ & $B$ & $N$  & \\
                       \hline
          1878 &14.80$\pm$0.39 &-20$\pm$19   &  43&1884 &14.27$\pm$0.07 &-1.7$\pm$1.2& 1138 & 1\\
          1890 &14.65$\pm$0.24 &-3.2$\pm$1.5 & 158&1893 &14.67$\pm$0.04 &-2.8$\pm$0.4 & 1817 & 1\\
          1902 &14.66$\pm$0.29 &-0.2$\pm$2.2 & 68&1906 &14.58$\pm$0.06 &-3.1$\pm$0.8 & 1096 & 1\\
          1986 &14.55$\pm$0.10 &-2.5$\pm$0.9 & 226&1989 &14.38$\pm$0.05 &-2.0$\pm$0.4 &2602  & 2\\
          1986 &14.55$\pm$0.08 &-3.4$\pm$0.8 & 297&1989 &14.38$\pm$0.04 &-2.6$\pm$0.3 & 3685 & 3\\
          1996 &14.47$\pm$0.14 & 0.3$\pm$1.6 & 158&2000 &14.49$\pm$0.05 &-2.8$\pm$0.4 &2113  & 2\\
          1996 &14.70$\pm$0.09 &-2.6$\pm$1.2 & 238&2000 &14.37$\pm$0.04 &-2.7$\pm$0.3 &3658  & 3\\
          2008 &14.63$\pm$0.27 &-1.0$\pm$1.9 &  66&2014 &14.46$\pm$0.05 &-1.7$\pm$0.7 &1467  & 2\\
          2008 &14.36$\pm$0.15 & 0.6$\pm$1.1 & 137&2014 &14.26$\pm$0.05 &-3.7$\pm$0.6 &3344  & 3\\
\hline
                       \end{tabular}\\
   \end{table}

\subsection{Correlation Between Solar Activity and Rotation}

\begin{figure}
   \includegraphics[width=0.45\textwidth]{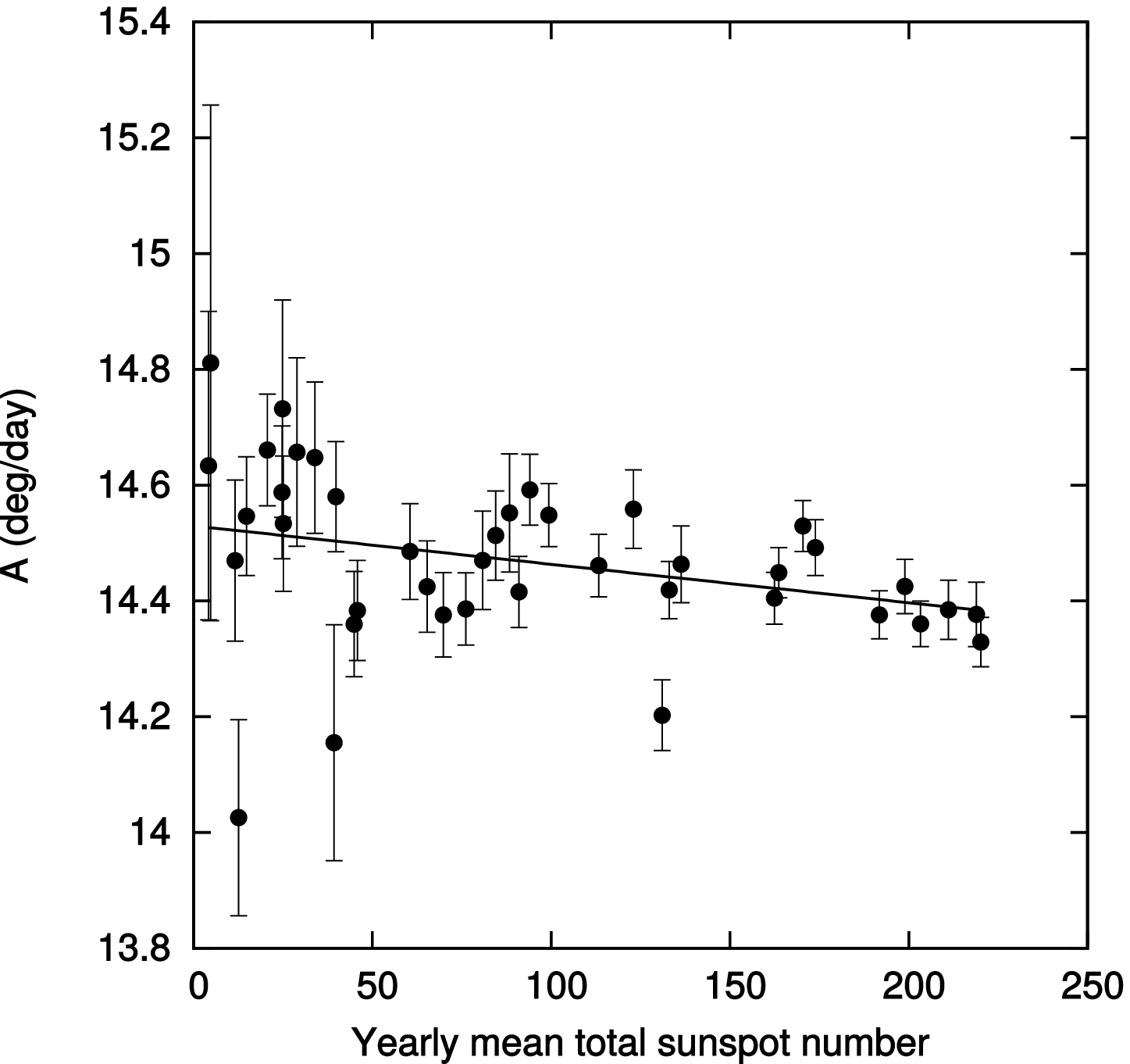}
   \includegraphics[width=0.45\textwidth]{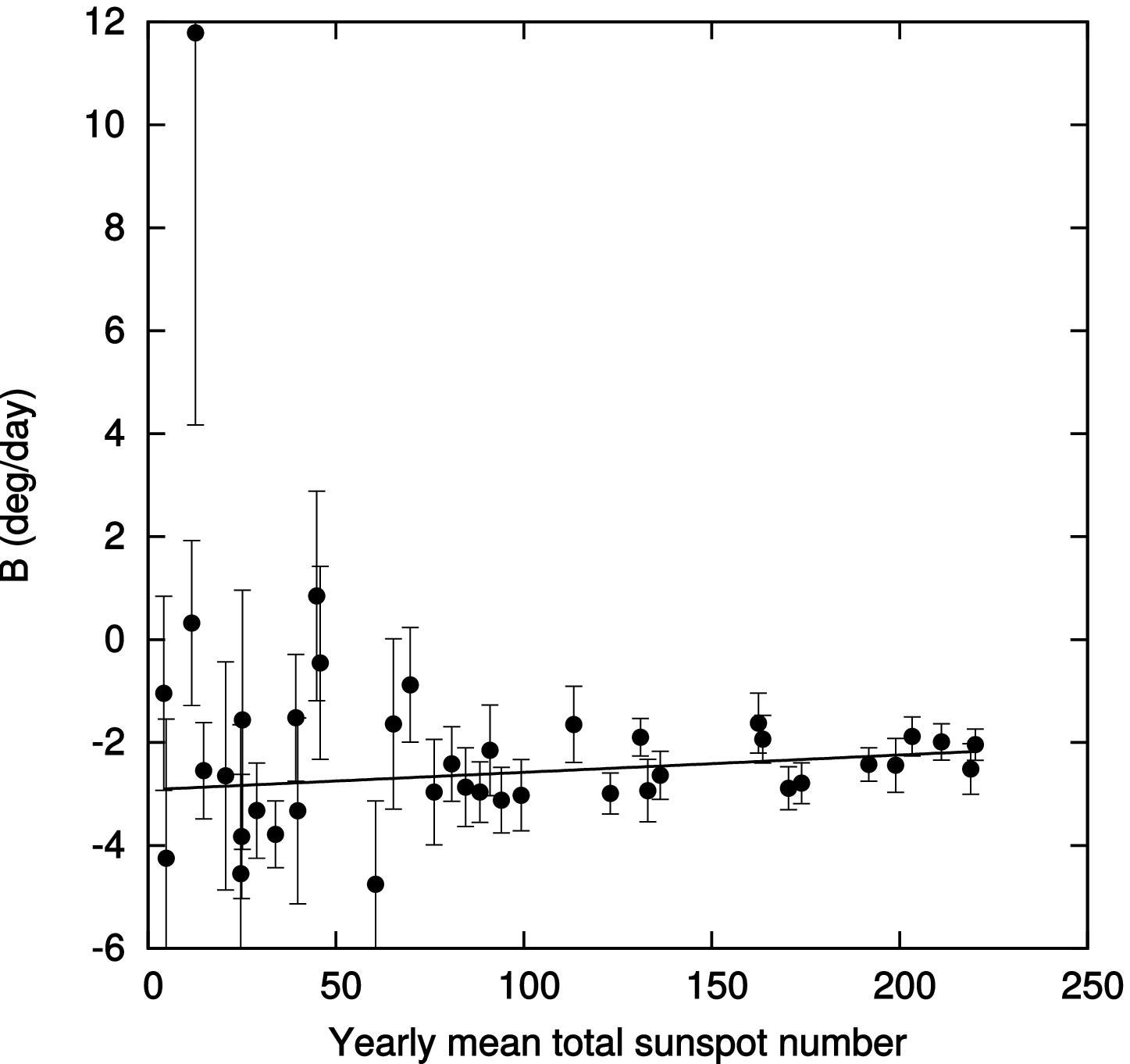}
      \caption{The differential-rotation parameters $A$ and $B$ for USAF/NOAA dataset shown {\it versus} yearly-mean 
total sunspot number. The lines represent least-squares fits using the Marquardt--Levenberg algorithm.  
              }
         \label{cornoaa}
   \end{figure}

In Figure \ref{cornoaa} the dependence of solar differential-rotation 
parameters on yearly mean total sunspot number is presented. The USAF/NOAA 
dataset, which has the highest correlation, is shown. Solid lines represent 
the least-square fits. To account for the errors of the solar differential-rotation 
parameters the data were fitted using an implementation of the 
least-squares Marquardt--Levenberg algorithm. The results of fits for the 
different datasets are summarized in Table \ref{fits}. As expected, the 
most significant values are obtained for the set with the highest correlation. 
The results, although not statistically significant for all datasets, show 
negative correlation between solar equatorial rotation rate and solar 
activity. The results for differential-rotation parameter $B$ are different 
for the various datasets. The GPR data result in an insignificant negative trend, while 
the data from USAF/NOAA and DPD result in a positive correlation 
with 2$\sigma$ statistical significance. Note that the values of $B$ are 
negative, so the positive correlation means a smaller absolute value during the 
maximum of activity. The results could imply that the correlation of 
differential-rotation parameter $B$ has changed.  In the period 1977\,--\,2016 
there is a positive correlation between differential-rotation parameter $B$ 
(significant on 2$\sigma$ level) and there is an (insignificant) negative 
correlation during 1874\,--\,1976. When the sets are combined the more numerous 
GPR dataset prevails.

  \begin{table}
      \caption[]{Least-square fits of yearly values of solar differential-rotation parameters 
{\it vs.} yearly-mean total sunspot numbers (YMTSN). 
(1 - GPR; 2 - USAF/NOAA; 3 - DPD; 4 - GPR+USAF/NOAA; 5 - GPR+DPD).}
         \label{fits}   
         \begin{tabular}{rrr}
            \hline                  
Dataset & slope $A$(YMTSN) [10$^{-3}\,^\circ$\,day$^{-1}$] & slope$B$(YMTSN) [10$^{-3}\,^\circ$\,day$^{-1}$] \\
\hline
      1 & -0.1$\pm$0.1 & -1.6$\pm$0.9  \\   
      2 & -0.6$\pm$0.2 &  3.0$\pm$1.5  \\
      3 & -0.2$\pm$0.2 &  3.4$\pm$1.7  \\
      4 & -0.3$\pm$0.1 & -0.6$\pm$0.8  \\
      5 & -0.2$\pm$0.1 & -0.2$\pm$0.8  \\
\hline
                       \end{tabular}\\
   \end{table}

\section{Discussion and Conclusion}

Our results show that the Sun rotates faster at low latitudes and more 
differentially at the minimum than at the maximum of activity.
This result is in agreement with the result of \citet{eddy1976} 
who measured solar surface rotation from sunspot drawings during 
Maunder minimum and found that solar equatorial rotation was 3\--\,4\,\% 
faster than today and that the differential rotation between Equator 
and latitudes of 20$^\circ$ was about three times larger.
However, \citet{abarbanell1981} using two original prints of Hevelius'
book, while \citet{eddy1976} used a copy, found that the solar rotation was
at the same level as today. The gradient of the differential rotation was 
found to be slightly steeper and not significantly different.  
More pronounced differential rotation during the Maunder minimum was also found
by \citet{ribes1993}.
Here, both secular and cycle-related changes of equatorial rotation rate 
that are anticorrelated with solar activity were 
found. This is not so clear for the differential-rotation parameter 
$B$ where no secular changes were found. 
The larger average rotation velocities in minimum than in other 
phases of cycle, {\it i.e.}, inverse correlation between solar 
equatorial-rotation rate and activity, was found by many authors 
using different methods and data 
\citep{lustig1983,gilman1984,balthasaretal1986,gupta1999,zuccarello2003,brajsa2006,jurdana2011,li2014,badalyan2016}. 

\citet{brun2004} numerically modelled the interaction of convection 
and rotation with magnetic field in deep spherical shells and found 
that the increase of magnetic field would result in the reduction of 
differential rotation. The reduction of differential rotation is due to 
the braking effect exerted by non-axisymmetric magnetic fields. This 
means that in the presence of strong magnetic fields net transport of 
angular momentum towards the Equator is less efficient. A similar result 
was obtained by \citet{lanza2006,lanza2007} by analytically solving the 
angular momentum transport equation within the convection zone of a rotating star. 
This provides the explanation of the observed effect. At the minimum of activity 
when the magnetic field is weaker, the angular momentum is transported 
more efficiently towards the Equator, which results in the observed increased 
equatorial rotation and more pronounced differential rotation at the minimum 
of activity.

The changes of the differential-rotation profile (differential-rotation 
parameter $B$) are much harder to detect than the changes of equatorial rotation. 
\citet{jurdana2011} analysed relationship between solar activity and
differential-rotation parameters obtained by tracing CBPs. They found 
inverse correlation of equatorial rotation rate with several times larger slope 
then obtained here and no significant correlation between parameter $B$ and activity. 
They attributed the lack of correlation for parameter $B$ to the errors  
of their data, which are more pronounced at high latitudes. 
The reduction of differential rotation in this work is detected 
(at the two-$\sigma$ level) in the 1977\,--\,2016 period, but not in the 1874\,--\,1976 epoch. 
This is most probably due to the errors, which are more pronounced at the beginning 
of the dataset where the amount of available data in years near minima is low. 
\citet{badalyan2016} examined the 22-year cycle of differential rotation 
and the rule of Genevyshev--Ohl by using the green coronal line brightness data from 1943 
onward and obtained a result consistent with inverse correlation of
both differential-rotation parameters and activity 
\citep[see Figures 3 and 4 in][]{badalyan2016}. Their finding of larger increases 
of equatorial rotation velocity between even and odd cycle than between 
odd and even is not visible in our data (see Figure \ref{aintime}). They 
did not find the difference between even and odd-cycle variation of 
differential-rotation parameter $B$ and their results show that 
the differential rotation is strongest just before maximum and 
weakest somewhere in between minimum and maximum  of activity. 
A similar result, although with much smaller statistical significance, 
is obtained in the GPR data. However, both amplitude and phase of 
the curve depend on the sampling, {\it i.e.}, how the phase of the cycle was chosen.

On examining the strength of solar maxima, it was noted that if the peak 
of equatorial rotation velocity is observed during minimum, then the next maximum 
is weaker than the preceding one.
The solar activity is high for 1950\,--\,2000, which can be regarded as the modern 
grand maximum of solar activity \citep{usoskin2017}. If the magnetic energy 
were to exceed about 20\,\% of the total kinetic energy, Maxwell stresses and 
magnetic torques may become strong enough to suppress the differential rotation 
almost entirely \citep[][and references therein]{brunetal2004}. 
This is not observed, so the Sun must have ways of avoiding this by expelling 
some of its magnetic flux. The observed stronger peaks of equatorial rotational 
velocity  might be a signature of such a process.
The most significant such event is observed during the minimum following the 
maximum of the Solar Cycle 19, which was the strongest one in the studied epoch. 
Cycle 20 is much weaker than the three previous and three following 
cycles, as if magnetic energy was expelled or consumed. No significant change 
of differential-rotation parameter $B$ can be observed accompanying the event.

Main result of our investigation is the finding that the Sun rotates more 
differentially at the minimum than at the maximum of activity during the 1977\,--\,2016 epoch. 
This is in agreement with theoretical predictions of reduced differential 
rotation in the presence of strong magnetic fields.
Inverse correlation between equatorial rotation and solar activity was found 
by many authors before and is corroborated here regardless of the recent revision 
of sunspot number. The secular decrease of rotation velocity accompanying the 
increase of activity stopped in the last part of the 20th century when solar activity 
started to decrease.
It was noted that when the significant peak of equatorial rotation velocity is 
observed during minimum of activity the strength of next maximum is smaller then 
the previous one. It was suggested that this finding might be connected to a decrease of 
the magnetic energy of the Sun.

\begin{acknowledgments}
The authors wish to thank Hubertus W\"ohl for useful sugestions and careful reading of the manuscript. We acknowledge the staff of Royal Observatory of 
Belgium, Brussels and the staff of Heliophysical Observatory, Debrecen, Hungary for 
maintaining and organizing the WDC-SILSO and DPD databases, respectively. 
This work was partly supported by the Croatian Science Foundation under the project 
6212 ''Solar and Stellar Variability", and in part by the University of Rijeka under 
project number 13.12.1.3.03. 
\end{acknowledgments}

\section*{Disclosure of Potential Conflicts of Interest}

The authors declare that they have no conflicts of interest.

\end{article} 

\end{document}